# Gate-Reconfigurable Single- and Double-Dot Transport in Trilayer $MoSe_2$

Seungwoo Lee[1,†], Minjun Park[1,†], Yunsang Noh[1,†], Sung Jin An[2,†], Soyun Kim[1], Minseo Cho[1], Dohun Kim[1], Takashi Taniguchi[3], Kenji Watanabe[4], Minkyung Jung[2,5,*], and Youngwook Kim[1,*]

[1]*Department of Physics and Chemistry, Daegu Gyeongbuk Institute of Science and Technology (DGIST), Daegu 42988, Republic of Korea*

[2]*DGIST Research Institute, Daegu Gyeongbuk Institute of Science and Technology (DGIST), Daegu 42988, Republic of Korea*

[3]*Research Center for Materials Nanoarchitectonics, National Institute for Materials Science, 1-1 Namiki, Tsukuba 305-0044, Japan*

[4]*Research Center for Electronic and Optical Materials, National Institute for Materials Science, 1-1 Namiki, Tsukuba 305-0044, Japan*

[5]*Department of Interdisciplinary Engineering, Daegu Gyeongbuk Institute of Science and Technology (DGIST), Daegu 42988, Republic of Korea*

[†]These authors contributed equally.

[*]E-mail: minkyung.jung@dgist.ac.kr, y.kim@dgist.ac.kr

**Abstract**

**We report gate-controlled quantum-dot transport in a trilayer $MoSe_2$ device that combines a graphite back gate beneath the active region, a separate global gate for conductive access regions, and local top finger gates. In the low-backgate regime, bias spectroscopy shows regular Coulomb-blockade diamonds characteristic of single-dot transport. As backgate is increased, additional low-bias structure develops beyond a simple single-dot pattern, indicating that the electrostatic landscape is reshaped and that a second dot becomes active in transport. In the higher-backgate regime, plunger-gate tuning and two-gate measurements establish a gate-reconfigurable double-dot configuration with two non-equivalent dots whose relative alignment and interdot coupling evolve with gate voltage. These results indicate that trilayer $MoSe_2$ supports electrically reconfigurable single- and double-dot transport in the present device architecture.**

## Introduction

Semiconductor quantum dots provide a versatile platform for studying confined charges and for developing spin-based quantum devices because electrostatic gates can define, tune, and probe their quantum states.[1,2] In established material systems such as GaAs[3], Si[4], and Ge[5], gate-defined dots have enabled Coulomb blockade measurements[6–8], charge sensing[9], spin initialization and readout[10], exchange control[11], and high-fidelity qubit operations.[12] This progress has established quantum-dot transport as a powerful framework for linking device electrostatics to the underlying electronic states.[10,11,13,14]

Transition-metal dichalcogenides (TMDCs) extend this framework to atomically thin semiconductors with an intrinsic band gap, reduced dimensionality, and electrically tunable carrier density.[15–18] Their strong spin-orbit coupling and valley-structured bands make them attractive for mesoscopic transport and electrically defined nanostructures.[19] Consistent with this promise, gate-defined dot transport, double-dot behavior, and charge detection have been demonstrated in $MoS_2$- and $WSe_2$-based devices, presenting that TMDCs can support controlled electrostatic confinement.[15,20–24]

Within this materials family, $MoSe_2$ has not yet been established to the same extent as $MoS_2$ and $WSe_2$ as a platform for gate-defined electronic quantum-dot transport. Recent low-density transport studies nevertheless indicate that high-quality $MoSe_2$ can support reliable electrical injection down to reduced carrier density[25–27], which makes it a timely material system for electrostatically defined nanostructures. $MoSe_2$ moreover offers a distinct combination of properties within the TMDC family: its conduction-band spin-orbit splitting (~20 meV) is roughly an order of magnitude larger than that of $MoS_2$[19,28,29], keeping the lowest conduction band spin-valley locked well above typical electron temperatures, and its large electron effective mass ($m^* \approx 0.8\ m_e$)[30] enhances the role of Coulomb interactions relative to the kinetic energy. The large mass is a double-edged property — it densifies the many-electron spectrum, but at low occupancy it opens access to interaction-dominated few-electron physics — making $MoSe_2$ a complementary rather than simply an additional platform. These considerations make $MoSe_2$ a natural platform for examining how quantum-dot transport develops under combined back-gate and local-gate control.

The device studied here combines a graphite back gate ($V_g$) beneath the active region, a separate global gate used to maintain conductive contact and access regions, and local top finger gates that shape the confinement potential. This architecture allows the broader electrostatic landscape and the local confinement to be tuned separately, which is particularly useful for tracing how the transport changes across different gate settings. Throughout this work, a gate-defined quantum dot denotes a charge island confined purely electrostatically by gate voltages within a continuous semiconductor channel; because the confinement is set by gate voltages alone, the dot configuration is reconfigurable in situ. Using this platform, we identify a low-$V_g$ regime dominated by single-dot transport and a higher-$V_g$ regime in which a second dot becomes active and a reconfigurable double-dot state emerges. Our focus is therefore not only to demonstrate quantum-dot transport in trilayer $MoSe_2$, but also to clarify how the combined graphite-back-gate and local-finger-gate architecture enables controlled evolution of the confinement landscape in this material.

## Results

High-quality $MoSe_2$ quantum-dot devices were fabricated by a Gel-Pak-based dry transfer process.[31,32] A graphite flake was first exfoliated onto a $SiO_2$/Si substrate, while an hBN flake was exfoliated onto a Gel-Pak stamp. The graphite flake was picked up with the hBN/Gel-Pak stamp and released onto the $SiO_2$/Si substrate to form the graphite/hBN stack. Contact-mode AFM scanning was then performed to remove transfer residue and to reduce bubbles and wrinkles at the graphite/hBN interface.[20,24,33] A $MoSe_2$ flake containing the selected trilayer region together with adjacent thicker regions was subsequently transferred onto the graphite/hBN stack, followed by a second contact-mode AFM scan. Finally, a top hBN flake was transferred onto the graphite/hBN/$MoSe_2$ stack. The top hBN was chosen to be smaller than the $MoSe_2$ flake so that a relatively large region of $MoSe_2$ remained unencapsulated for subsequent top-contact fabrication, as depicted in Fig. 1a and Fig. 1c.

We used Sb/Au top contacts, motivated by recent demonstrations that semimetal contacts suppress metal-induced gap states and Fermi-level pinning, providing ohmic injection into Mo-based TMDCs at low carrier density[34–38], as shown in Fig. 1c. Sb, which forms the interface with $MoSe_2$ beneath the Au capping layer, was chosen over Bi for its higher melting point and better process stability[39]; conventional Cr/Au and Ti/Au contacts did not yield measurable conduction in our devices at low carrier density. The thicker $MoSe_2$ regions outside the selected trilayer active area were then removed by etching. Further details of device fabrication are provided in the Methods section and the Supplementary Information. The finger gates and the graphite back-gate contact were subsequently defined using Cr/Au, as displayed in Fig. 1b and

Fig. 1d. Here, $V_g$ denotes the voltage applied to the graphite back gate beneath the active region, and $V_{gg}$ the global gate voltage applied to the doped Si substrate, which acts through the $SiO_2$ layer on the regions outside the graphite back gate and keeps the contact and access regions conductive. $V_P$ denotes the plunger-gate voltage, and $V_L$, $V_M$, and $V_R$ the left, middle, and right constriction-gate voltages, respectively.

We next characterized the basic transport properties of the device before entering the quantum-dot regime. Figure 1e presents a typical semiconducting transfer characteristic behavior measured as a function of graphite back-gate voltage $V_g$ while the global gate $V_{gg}$ was fixed at 70 V. The transfer curve exhibits an on/off ratio on the order of $10^4$, confirming that the graphite back gate effectively tunes the carrier density in the active region, while the global gate keeps the contact and access regions conductive during the measurement. Figure 1f displays a current-voltage characteristic measured at $V_g$ = 2 V, indicating that the top contacts are well formed and provide ohmic carrier injection. These measurements establish the basic gate response and contact quality of the present device architecture and provide the starting point for the quantum-dot measurements discussed below.

Having established the basic transport characteristics of the device, we now turn to the low-$V_g$ regime. Figure 2 displays the corresponding bias spectroscopy. With $V_{gg}$ = 70 V, $V_g$ = 0.72 V in Fig. 2a and 0.73 V in Fig. 2b, $V_M$ = -1.0 V, and $V_L$ = $V_R$ = -1.3 V, the bias spectroscopy as a function of the plunger gate $V_P$ exhibits a regular series of Coulomb-blockade diamonds with nearly periodic spacing in gate voltage. The two panels show closely similar diamond patterns, with a slight shift of the resonances toward lower $V_P$ as $V_g$ is increased by 0.01 V. In this window, the low-bias transport exhibits single-dot behavior, in the sense that transport is dominated mainly by single dot over most of the measured gate voltage window. At the same time, the present device remains in the many-electron regime rather than in a clearly resolved few-electron limit. Several larger diamond features already appear below about 1.35 V in Supplementary Note S3, but they are not sufficiently resolved for a reliable few-electron analysis.

To quantify this low-$V_g$ regime, we extracted the addition energy $E_{add}$ — the energy required to add one electron to the dot — and a plunger-gate lever arm from the representative diamonds in Fig. 2a. From the characteristic diamond extent in bias, we estimate $\Delta V_{SD} \approx 2.1$ mV, which gives $\underline{E}_{add} \approx e\Delta V_{SD} \approx 2.1$ meV. To determine the plunger-gate spacing, we take a line cut at $V_{SD}$ = 133 μV and identify Coulomb peak positions at $V_P$ = 1.4169, 1.4351, 1.4555, 1.4742, and 1.4945 V. These values give an average peak spacing of $\Delta V_P \approx 18.7$ mV. Using the relation $E_{add} = \alpha_P e\Delta V_P$, we obtain a plunger-gate lever arm of $\alpha_P \approx 0.11$. In addition, fits of representative Coulomb peaks to a thermally broadened lineshape yield an electron temperature of $T_e \approx 2.3$ K. These values provide a quantitative characterization of the low-$V_g$ regime.

The in-plane magnetic-field dependence was investigated under the same low-$V_g$ conditions as in Fig. 2a, with $V_g = 0.72$ V. Figures 2c and 2d present the corresponding Coulomb-diamond patterns measured at $B_{\|} = 1$ T and 5 T, respectively. Both finite-field datasets remain very similar to the zero-field result in Fig. 2a, with no clear field-induced shift, splitting, or systematic reconstruction over the measured field range. This weak field dependence is consistent with the present device still being outside a clearly resolved few-electron regime. Because $MoSe_2$ has a relatively large effective mass, the many-electron spectrum can become dense and the level spacing correspondingly small, so that field-induced corrections are not easily resolved in direct transport. The orbital effect of an in-plane field is also expected to be weak in this atomically thin geometry. Under these conditions, the low-$V_g$ response is governed mainly by electrostatic confinement and charging, while any magnetic-field-induced changes remain smaller than the linewidth and energy resolution of the present data.

We now examine how the low-$V_g$ transport signature changes as the graphite back-gate voltage is increased. Figure 3 summarizes this change under otherwise comparable gate settings. Here, $V_{gg}$, $V_M$, $V_L$, and $V_R$ are fixed, $B_{\|} = 1$ T, and only $V_g$ is varied from 1.0 V in Fig. 3a to 1.1 V in Fig. 3b. At $V_g = 1.0$ V, the bias spectroscopy remains close to a single-dot limit, although the diamonds are already less regular than in the lower-$V_g$ regime of Fig. 2. As $V_g$ is increased further, additional low-bias structure develops and the spectroscopy is no longer captured by a simple single-dot picture. The higher-$V_g$ data contain features that are difficult to account for by one dot alone.

This reconfiguration occurs even though the local-gate settings remain essentially unchanged, which identifies the graphite back gate as the main control parameter. The data indicate that the effective confinement landscape already contains two non-equivalent dots. At lower $V_g$, transport is dominated by the lower dot, so the spectroscopy remains close to a single-dot limit over the accessible gate window. As $V_g$ is increased, the second dot becomes active in transport and the spectroscopy develops into a double-dot configuration. The change from Fig. 3a to Fig. 3b is therefore understood as a back-gate-driven evolution of a non-equivalent double-well potential rather than as a consequence of a different plunger-gate condition. Although the asymmetry between the two dots may reflect slight geometric variations of the lithographically defined gates together with residual potential inhomogeneity, the reproducible evolution with $V_g$ and $V_P$ shows that the resulting double-dot response is systematically controlled by the applied gates.

This picture also provides a simple description of the higher-$V_g$ regime through the schematics in Fig. 4a and Fig. 4b. In Fig. 4a, the electrostatic landscape is represented by an asymmetric double-well potential containing two non-equivalent dots. Within this picture, the graphite back gate mainly shifts the overall electrostatic landscape and brings the second dot into the transport window as $V_g$ is increased, while the plunger gate acts within the same double-well potential. As illustrated schematically in Fig. 4b, increasing $V_P$ lowers the central barrier and strengthens the coupling between the two dots, so that the transport evolves from a more separated double-dot state toward a more merged single-dot state. This interpretation agrees with the bias spectroscopy in Fig. 4c. At lower $V_P$, the pattern exhibits multi-dot behavior, whereas at higher $V_P$ it moves toward a more merged single-dot state. Between about $V_P$ = 0.88 V and 0.98 V, the data pass through an intermediate regime in which both characters are present.

The same trend is observed in the current maps measured as a function of $V_L$ and $V_R$ (Fig. 4d-4f). At $V_P$ = 0.75 V in Fig. 4d, the pattern is consistent with a well-separated double-dot configuration, with more than one slope family visible in the conductance features. At $V_P$ = 0.95 V in Fig. 4e, the map takes an intermediate form, consistent with partial merging of the two dots. At $V_P$ = 1.10 V in Fig. 4f, the pattern becomes more single-dot-like, with a more unified gate response. This systematic progression with $V_P$ supports the view that, once the back gate has brought the device into the higher-$V_g$ regime, the plunger gate can further tune the central barrier and thereby modify the interdot coupling.

To quantify the double-dot regime more directly, we focus on the dashed rectangular region of Fig. 4d, which is magnified in Fig. 5a. The honeycomb-like charge-stability pattern and the finite-bias triangles measured at point A in Fig. 5b are consistent with a capacitively coupled double quantum dot. From the gate periods $\Delta V_L$ and $\Delta V_R$ in Fig. 5a, we estimate the effective gate capacitances between $V_L$ and QD1, and between $V_R$ and QD2, to be $C_L = e/\Delta V_L \approx 1.26$ aF and $C_R = e/\Delta V_R \approx 0.44$ aF, respectively. Using the bias-triangle extents $\delta V_L$ and $\delta V_R$ at $V_{SD}$ = 1.5 mV, we further estimate lever arms $\alpha_L = V_{SD}/\delta V_L \approx 0.031$ and $\alpha_R = V_{SD}/\delta V_R \approx 0.0125$. Within a standard local capacitance model[40], these values give effective total dot capacitances of $C_1 = C_L/\alpha_L \approx 40.6$ aF and $C_2 = C_R/\alpha_R \approx 35.1$ aF, a mutual capacitance of $C_m \approx 12.7$ aF, and charging energies of $E_C^L \approx 4.45$ meV and $E_C^R \approx 5.15$ meV. Because this analysis is restricted to the selected local configuration of Fig. 4d, and because the confinement landscape evolves with gate voltage, these parameters should be regarded as local effective values rather than unique device constants. Nevertheless, they support the interpretation of this region as a finite and appreciably coupled double-dot configuration.

**Discussion**

The main implication of these measurements is that trilayer $MoSe_2$ can support not only gate-defined single-dot transport, but also controlled reconfiguration into a double-dot state within the same device architecture. In the present structure, the graphite back gate determines the broader electrostatic landscape and controls whether one or two dots participate in transport, whereas the local gates tune their relative alignment and interdot coupling once the double-dot regime is reached.

While the present device successfully demonstrates gate-reconfigurable dot transport, it currently operates in the many-electron regime. A primary motivation for exploring quantum dots in transition-metal dichalcogenides is to harness their strong spin-orbit coupling and valley-structured bands. However, extracting precise spin states, valley degrees of freedom, and Landé g-factors requires resolving the Zeeman splitting of individual energy levels. As observed in our in-plane magnetic-field measurements, the relatively large effective mass of $MoSe_2$ leads to a dense many-electron spectrum with correspondingly small level spacing, making field-induced corrections difficult to resolve in direct transport.

Therefore, transitioning from this many-electron state to a clearly resolved few-electron limit remains a critical next step to fully exploit the unique properties of $MoSe_2$. A more reliable approach toward this lowest-occupancy limit will require a cleaner heterostructure, stronger electrostatic confinement, and improved contact and access-region design. By reducing the effective dot size and sharpening the gate-defined potential, this platform should allow sharper extraction of interdot charging and coupling parameters, ultimately opening the way to more detailed spin and valley spectroscopy of coupled dots in TMDCs.

## Methods

### Device fabrication

We fabricated graphite/hBN/trilayer $MoSe_2$/hBN heterostructures by a dry pick-up method using Gel-Pak stamps. A bottom hBN flake, about 17.5 nm thick, was used to pick up a graphite flake and transferred onto a $SiO_2$/Si substrate. The stack was then annealed at 600 °C for 4 h in a continuously flowing Ar/H2 atmosphere containing 3% H2. A trilayer $MoSe_2$ flake and then a top hBN flake, about 13.3 nm thick, were sequentially transferred onto the bottom graphite/hBN structure, with contact-mode AFM cleaning performed after each transfer step. All contact-mode cleaning was carried out using PPP-NCHR probes (NANOSENSORS, nominal spring constant 42 N/m) at normal forces below 50 nN with scan speed of 0.1 ~ 0.5 Hz.

The top hBN was intentionally made smaller than the $MoSe_2$ flake to leave an unencapsulated region for direct top contacting. Because edge-contact test structures did not provide reliable electrical conduction, Sb/Au top contacts, 50 nm and 20 nm, were fabricated on the unencapsulated $MoSe_2$ region by electron-beam lithography and electron-beam evaporation, followed by an additional contact-mode AFM cleaning step. The exposed bulk $MoSe_2$ was then removed by two-step reactive ion etching using $CF_4$ plasma (40 sccm, 40 W) and $O_2$ plasma (50 sccm, 20 W). Local finger gates, 3 nm Cr and 20 nm Au, and connecting electrodes, 5 nm Cr and 95 nm Au, were subsequently fabricated by electron-beam lithography and electron-beam evaporation. Further fabrication details are provided in Supplementary Note S1.

### Transport measurement

Electrical transport measurements were performed in an attocube attoDRY 2100 cryostat at a base temperature of 2 K. Transport data were acquired with a DC bias voltage applied by a Basel DAC SP 927, while the resulting current was amplified by a Basel SP 983c IV converter and measured with a Keysight 34460A digital multimeter.

### Extraction of addition energy, plunger-gate lever arm, and electron temperature

The addition-energy scale and the plunger-gate lever arm were extracted from representative Coulomb diamonds measured in the low-$V_g$ regime of Fig. 2a. The addition-energy scale was estimated from the characteristic diamond extent along the source-drain bias direction, using the relation $E_{add} \approx e\Delta V_{SD}$ where $\Delta V_{SD}$ denotes the full diamond height in bias.

To determine the plunger-gate spacing, a line cut was taken at fixed bias, $V_{SD}$ = 133 μV, and the Coulomb peak positions were identified from the resulting current trace as a function of $V_P$. The average peak spacing was then obtained from $\Delta V_P = \langle V_{P,i+1} - V_{P,i} \rangle$. The plunger-gate lever arm $\alpha_P$ was estimated from the relation $E_{add} = \alpha_P \times e\Delta V_P$ which gives $\alpha_P = E_{add} / (e\Delta V_P)$.

To estimate the electron temperature, representative Coulomb peaks were fitted with a thermally broadened lineshape of the form $I(V_P) = I_{off} + A \cosh^{-2}[\alpha_P \times e(V_P - V_0) / (2k_B Te)]$ where $I_{off}$ is an offset current, A is the peak amplitude, $V_0$ is the peak center, and $T_e$ is the electron temperature. The extracted peak widths were further checked using the standard thermal-broadening relation FWHM ≈ $3.53\ k_B Te / (\alpha_P \times e)$.

The extracted $E_{add}$ is an addition-energy scale that in general contains both the electrostatic charging energy and any single-particle level spacing; because individual excited states are not resolved in the present many-electron regime, these contributions are not separated.

### Extraction of normalized triple-point splitting in the double-dot regime

For the double-dot analysis, parameters were extracted from the honeycomb lattice in Fig. 5(a) and the bias triangles in Fig. 5(b). The lever arms (α) were determined using the relation $\alpha = V_{SD} / \delta V$, where $\delta V$ denotes the gate voltage extension of the bias triangle at an applied source-drain bias $V_{SD}$. Individual gate capacitances ($C_g$) were calculated from the gate period ($\Delta V_g$) via $C_g = e / \Delta V_g$, and the total capacitance ($C_{tot}$) of each dot was obtained through $C_{tot} = C_g / \alpha$. The mutual capacitance ($C_m$) was estimated from the triple-point splitting ($\Delta V_g^m$) using the relation $C_m = (\Delta V_L^m / \Delta V_R^m)\ C_2$. Correspondingly, the charging energies of the left and right dots ($E_C^L$, $E_C^R$), which account for the interdot electrostatic coupling, were calculated using the equation $E_C^{L(R)} = (e^2 / C_{1(2)})\ \{1 - Cm^2 / (C_1 C_2)\}^{-1}$. The normalized splitting, defined as the ratio of the triple-point splitting to the gate period, was used as a local phenomenological measure of the coupling strength. These parameters characterize the specific gate configuration, as the applied gate voltages influence not only the charge occupancy but also the confinement potential and interdot barriers.

## References

[1] G. Burkard, T.D. Ladd, A. Pan, J.M. Nichol, and J.R. Petta, "Semiconductor spin qubits," *Rev. Mod. Phys.* **95**, 025003 (2023).

[2] D. Loss, and D.P. DiVincenzo, "Quantum computation with quantum dots," *Phys. Rev. A* **57**, 120–126 (1998).

[3] R. Hanson, L.P. Kouwenhoven, J.R. Petta, S. Tarucha, and L.M.K. Vandersypen, "Spins in few-electron quantum dots," *Rev. Mod. Phys.* **79**, 1217–1265 (2007).

[4] F.A. Zwanenburg, A.S. Dzurak, A. Morello, M.Y. Simmons, L.C.L. Hollenberg, G. Klimeck, S. Rogge, S.N. Coppersmith, and M.A. Eriksson, "Silicon quantum electronics," *Rev. Mod. Phys.* **85**, 961–1019 (2013).

[5] G. Scappucci, C. Kloeffel, F.A. Zwanenburg, D. Loss, M. Myronov, J.-J. Zhang, S. De Franceschi, G. Katsaros, and M. Veldhorst, "The germanium quantum information route," *Nat. Rev. Mater.* **6**, 926–943 (2021).

[6] U. Meirav, M.A. Kastner, and S.J. Wind, "Single-electron charging and periodic conductance resonances in GaAs nanostructures," *Phys. Rev. Lett.* **65**, 771–774 (1990).

[7] M.A. Kastner, "The single-electron transistor," *Rev. Mod. Phys.* **64**, 849–858 (1992).

[8] H. van Houten, "Coulomb blockade oscillations in semiconductor nanostructures," *Surf. Sci.* **263**, 442–445 (1992).

[9] M. Field, C.G. Smith, M. Pepper, D.A. Ritchie, J.E.F. Frost, G.A.C. Jones, and D.G. Hasko, "Measurements of Coulomb blockade with a noninvasive voltage probe," *Phys. Rev. Lett.* **70**, 1311–1314 (1993).

[10] J.M. Elzerman, R. Hanson, L.H. Willems van Beveren, B. Witkamp, L.M.K. Vandersypen, and L.P. Kouwenhoven, "Single-shot read-out of an individual electron spin in a quantum dot," *Nature* **430**, 431–435 (2004).

[11] J.R. Petta, A.C. Johnson, J.M. Taylor, E.A. Laird, A. Yacoby, M.D. Lukin, C.M. Marcus, M.P. Hanson, and A.C. Gossard, "Coherent Manipulation of Coupled Electron Spins in Semiconductor Quantum Dots," *Science* **309**, 2180–2184 (2005).

[12] X. Xue, M. Russ, N. Samkharadze, B. Undseth, A. Sammak, G. Scappucci, and L.M.K. Vandersypen, "Quantum logic with spin qubits crossing the surface code threshold," *Nature* **601**, 343–347 (2022).

[13] A. Noiri, K. Takeda, T. Nakajima, T. Kobayashi, A. Sammak, G. Scappucci, and S. Tarucha, "Fast universal quantum gate above the fault-tolerance threshold in silicon," *Nature* **601**, 338–342 (2022).

[14] N.W. Hendrickx, W.I.L. Lawrie, M. Russ, F. van Riggelen, S.L. de Snoo, R.N. Schouten, A. Sammak, G. Scappucci, and M. Veldhorst, "A four-qubit germanium quantum processor," *Nature* **591**, 580–585 (2021).

[15] K. Wang, K. De Greve, L.A. Jauregui, A. Sushko, A. High, Y. Zhou, G. Scuri, T. Taniguchi, K. Watanabe, M.D. Lukin, H. Park, and P. Kim, "Electrical control of charged carriers and excitons in atomically thin materials," *Nat. Nanotechnol.* **13**, 128–132 (2018).

[16] M. Chhowalla, H.S. Shin, G. Eda, L.-J. Li, K.P. Loh, and H. Zhang, "The chemistry of two-dimensional layered transition metal dichalcogenide nanosheets," *Nat. Chem.* **5**, 263–275 (2013).

[17] K.F. Mak, C. Lee, J. Hone, J. Shan, and T.F. Heinz, "Atomically Thin $MoS_2$: A New Direct-Gap Semiconductor," *Phys. Rev. Lett.* **105**, 136805 (2010).

[18] B. Radisavljevic, A. Radenovic, J. Brivio, V. Giacometti, and A. Kis, "Single-layer $MoS_2$ transistors," *Nat. Nanotechnol.* **6**, 147–150 (2011).

[19] D. Xiao, G.-B. Liu, W. Feng, X. Xu, and W. Yao, "Coupled Spin and Valley Physics in Monolayers of $MoS_2$ and Other Group-VI Dichalcogenides," *Phys. Rev. Lett.* **108**, 196802 (2012).

[20] J. Boddison-Chouinard, A. Bogan, N. Fong, P. Barrios, J. Lapointe, K. Watanabe, T. Taniguchi, A. Luican-Mayer, and L. Gaudreau, "Charge Detection Using a van der Waals Heterostructure Based on Monolayer $WSe_2$," *Phys. Rev. Appl.* **18**, 054017 (2022).

[21] Z.-Z. Zhang, X.-X. Song, G. Luo, G.-W. Deng, V. Mosallanejad, T. Taniguchi, K. Watanabe, H.-O. Li, G. Cao, G.-C. Guo, F. Nori, and G.-P. Guo, "Electrotunable artificial molecules based on van der Waals heterostructures," *Sci. Adv.* **3**, e1701699 (2017).

[22] R. Pisoni, Z. Lei, P. Back, M. Eich, H. Overweg, Y. Lee, K. Watanabe, T. Taniguchi, T. Ihn, and K. Ensslin, "Gate-tunable quantum dot in a high quality single layer $MoS_2$ van der Waals heterostructure," *Appl. Phys. Lett.* **112**, 123101 (2018).

[23] S. Davari, J. Stacy, A.M. Mercado, J.D. Tull, R. Basnet, K. Pandey, K. Watanabe, T. Taniguchi, J. Hu, and H.O.H. Churchill, "Gate-Defined Accumulation-Mode Quantum Dots in Monolayer and Bilayer $WSe_2$," *Phys. Rev. Appl.* **13**, 054058 (2020).

[24] J. Boddison-Chouinard, A. Bogan, N. Fong, K. Watanabe, T. Taniguchi, S. Studenikin, A. Sachrajda, M. Korkusinski, A. Altintas, M. Bieniek, P. Hawrylak, A. Luican-Mayer, and L. Gaudreau, "Gate-controlled quantum dots in monolayer $WSe_2$," *Appl. Phys. Lett.* **119**, 133104 (2021).

[25] S. Larentis, B. Fallahazad, and E. Tutuc, "Field-effect transistors and intrinsic mobility in

ultra-thin $MoSe_2$ layers," *Appl. Phys. Lett.* **101**, 223104 (2012).

[26] B. Chamlagain, Q. Li, N.J. Ghimire, H.-J. Chuang, M.M. Perera, H. Tu, Y. Xu, M. Pan, D. Xaio, J. Yan, D. Mandrus, and Z. Zhou, "Mobility Improvement and Temperature Dependence in $MoSe_2$ Field-Effect Transistors on Parylene-C Substrate," *ACS Nano* **8**, 5079–5088 (2014).

[27] A. Allain, J. Kang, K. Banerjee, and A. Kis, "Electrical contacts to two-dimensional semiconductors," *Nat. Mater.* **14**, 1195–1205 (2015).

[28] A. Kormányos, G. Burkard, M. Gmitra, J. Fabian, V. Zólyomi, N.D. Drummond, and V. Fal'ko, "k·p theory for two-dimensional transition metal dichalcogenide semiconductors," *2D Mater.* **2**, 022001 (2015).

[29] K. Kośmider, J.W. González, and J. Fernández-Rossier, "Large spin splitting in the conduction band of transition metal dichalcogenide monolayers," *Phys. Rev. B* **88**, 245436 (2013).

[30] S. Larentis, H.C.P. Movva, B. Fallahazad, K. Kim, A. Behroozi, T. Taniguchi, K. Watanabe, S.K. Banerjee, and E. Tutuc, "Large effective mass and interaction-enhanced Zeeman splitting of $K$-valley electrons in $MoSe_2$," *Phys. Rev. B* **97**, 201407 (2018).

[31] Y. Kim, A.C. Balram, T. Taniguchi, K. Watanabe, J.K. Jain, and J.H. Smet, "Even denominator fractional quantum Hall states in higher Landau levels of graphene," *Nat. Phys.* **15**, 154–158 (2019).

[32] A. Castellanos-Gomez, M. Buscema, R. Molenaar, V. Singh, L. Janssen, H.S.J. van der Zant, and G.A. Steele, "Deterministic transfer of two-dimensional materials by all-dry viscoelastic stamping," *2D Mater.* **1**, 011002 (2014).

[33] Y. Kim, P. Herlinger, T. Taniguchi, K. Watanabe, and J.H. Smet, "Reliable Postprocessing Improvement of van der Waals Heterostructures," *ACS Nano* **13**, 14182–14190 (2019).

[34] P.-C. Shen, C. Su, Y. Lin, A.-S. Chou, C.-C. Cheng, J.-H. Park, M.-H. Chiu, A.-Y. Lu, H.-L. Tang, M.M. Tavakoli, G. Pitner, X. Ji, Z. Cai, N. Mao, J. Wang, V. Tung, J. Li, J. Bokor, A. Zettl, C.-I. Wu, T. Palacios, L.-J. Li, and J. Kong, "Ultralow contact resistance between semimetal and monolayer semiconductors," *Nature* **593**, 211–217 (2021).

[35] Y. Wang, J.C. Kim, R.J. Wu, J. Martinez, X. Song, J. Yang, F. Zhao, A. Mkhoyan, H.Y. Jeong, and M. Chhowalla, "Van der Waals contacts between three-dimensional metals and two-dimensional semiconductors," *Nature* **568**, 70–74 (2019).

[36] W. Li, X. Gong, Z. Yu, L. Ma, W. Sun, S. Gao, Ç. Köroğlu, W. Wang, L. Liu, T. Li, H. Ning, D. Fan, Y. Xu, X. Tu, T. Xu, L. Sun, W. Wang, J. Lu, Z. Ni, J. Li, X. Duan, P. Wang, Y.

Nie, H. Qiu, Y. Shi, E. Pop, J. Wang, and X. Wang, "Approaching the quantum limit in two-dimensional semiconductor contacts," *Nature* **613**, 274–279 (2023).

[37] T.D. Ngo, T. Huynh, H. Jung, F. Ali, J. Jeon, M.S. Choi, and W.J. Yoo, "Modulation of Contact Resistance of Dual-Gated $MoS_2$ FETs Using Fermi-Level Pinning-Free Antimony Semi-Metal Contacts," *Adv. Sci.* **10**, 2301400 (2023).

[38] T. Su, Y. Li, Q. Wang, W. Zhao, L. Cao, and Y.S. Ang, "Semimetal contacts to monolayer semiconductor: weak metalization as an effective mechanism to Schottky barrier lowering," *J. Phys. Appl. Phys.* **56**, 234001 (2023).

[39] A.-S. Chou, T. Wu, C.-C. Cheng, S.-S. Zhan, I.-C. Ni, S.-Y. Wang, Y.-C. Chang, S.-L. Liew, E. Chen, W.-H. Chang, C.-I. Wu, J. Cai, H.-S.P. Wong, and H. Wang, "Antimony Semimetal Contact with Enhanced Thermal Stability for High Performance 2D Electronics," in *2021 IEEE Int. Electron Devices Meet. IEDM*, (2021), p. 7.2.1-7.2.4.

[40] W.G. van der Wiel, S. De Franceschi, J.M. Elzerman, T. Fujisawa, S. Tarucha, and L.P. Kouwenhoven, "Electron transport through double quantum dots," *Rev. Mod. Phys.* **75**, 1–22 (2002).

## Acknowledgement

We thank Klaus Ensslin and Thomas Ihn for helpful discussions. The work from DGIST was supported by the National Research Foundation of Korea (NRF) (Grant No. RS-2025-00557717, RS-2023-00269616, RS-2025-02315685) and the Nano and Material Technology Development Program through the National Research Foundation of Korea (NRF) funded by Ministry of Science and ICT (No. RS-2024-00444725). We also acknowledge the partner group program of the Max Planck Society. Part of this work was supported by Global Partnership Program of Leading Universities in Quantum Science and Technology (RS-2025-02317602). M.J. acknowledges support from the National Research Foundation of Korea (NRF) (Grant No. RS-2023-NR076585, RS-2025-25463923), the DGIST R&D Program of the Ministry of Science, ICT, and Future Planning (26-ET-02). K.W. and T.T. acknowledge support from the JSPS KAKENHI (Grant Numbers 21H05233 and 23H02052), the CREST (JPMJCR24A5), JST and World Premier International Research Center Initiative (WPI), MEXT, Japan.

## Author Contributions

S. L, M. J, and Y. K conceived the project. S. L, Y. N, M. P, S. K, M. C., and D. K carried out the device fabrication. The low-temperature measurements were done by S. L, S.J. A, Y. N, M. P. under supervision of M. J, and Y. K. Datasets were analyzed by M. P, S. L, M. J, and Y. K. T. T and K. W synthesized the h-BN crystals. All authors contributed to the manuscript writing.

## Competing Interests

The authors declare no competing interest

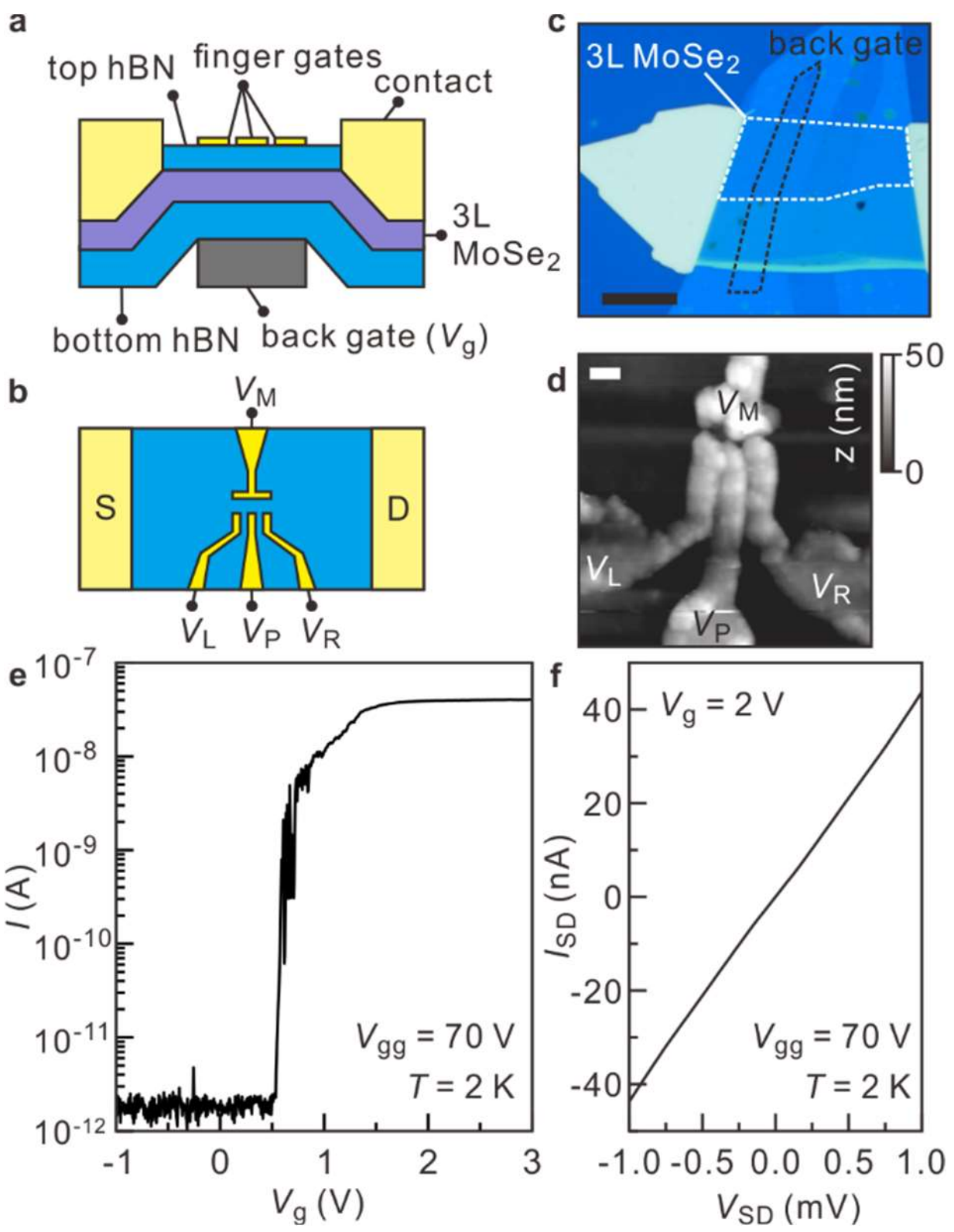

**Figure 1| Device structure and basic electrical characterization of the trilayer $MoSe_2$ quantum-dot device.** (a) Schematic cross section of the device, consisting of trilayer $MoSe_2$ encapsulated by hBN, local top finger gates, source and drain contacts, and a graphite back gate ($V_g$) beneath the active region. (b) Schematic top-view layout of the gate-defined constriction region between source (S) and drain (D). Here, $V_P$ denotes the plunger gate, while $V_L$, $V_M$, and $V_R$ denote the left, middle, and right constriction gates. (c) Optical image of the device after Sb contact fabrication and before etching. White dotted line indicates trilayer $MoSe_2$ area. At this stage, thicker $MoSe_2$ regions are still present and have not yet been removed. The highlighted regions indicate the trilayer $MoSe_2$ area selected as the active device region and the underlying graphite back gate. The scale bar is 10 μm. (d) AFM topography image of the central active region after device fabrication, showing the local gate geometry around the constricted channel. The scale bar is 100 nm. (e) Transfer characteristic measured as a function of graphite back-gate voltage at $V_{gg}$ = 70 V and $T$ = 2 K. (f) Current-bias characteristic measured at $V_{gg}$ = 70 V and $T$ = 2 K.

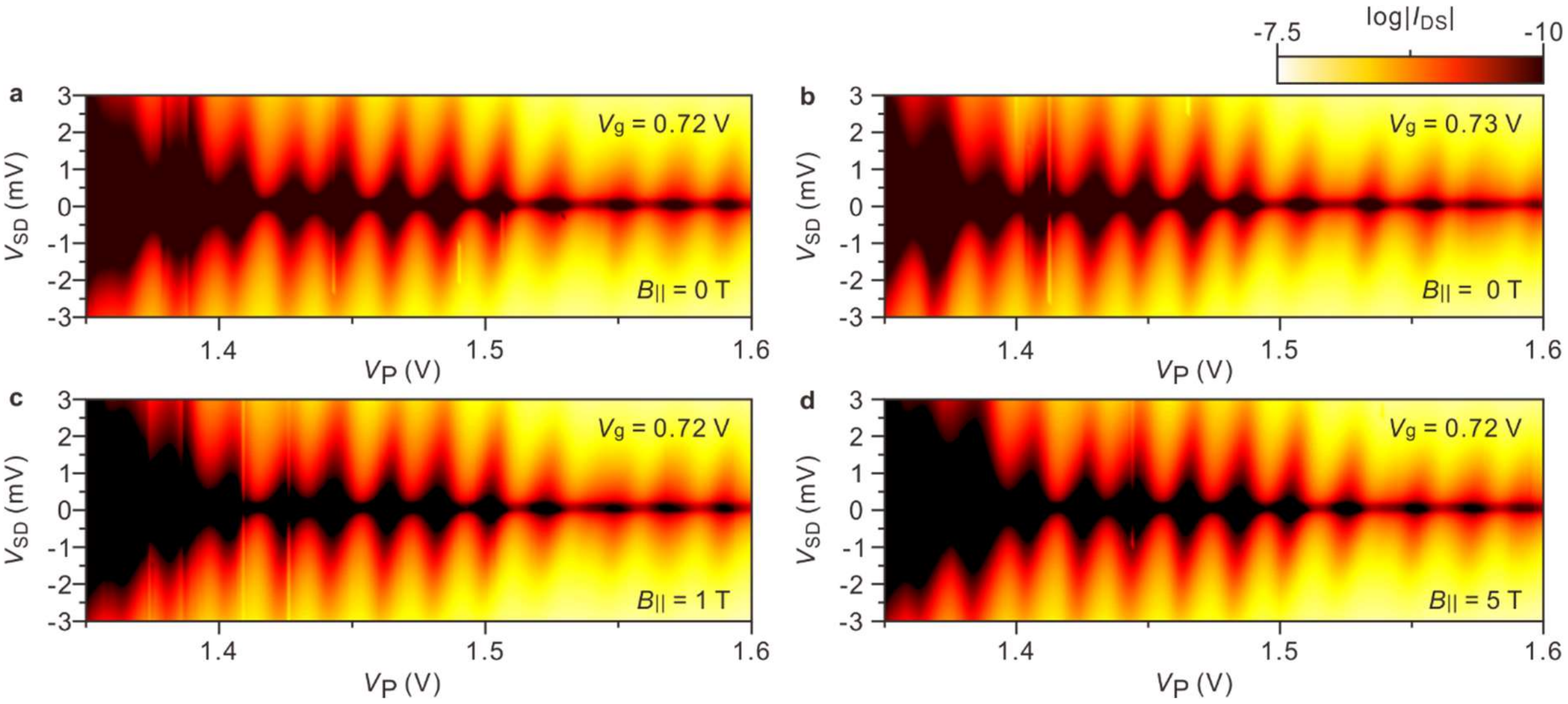

**Figure 2| Low-gate bias spectroscopy and its weak in-plane magnetic-field dependence.** Bias spectroscopy as a function of plunger-gate voltage $V_P$, measured at $V_{gg}$ = 70 V, $V_M$ = -1.0 V, and $V_L$ = $V_R$ = -1.3 V, with (a) $V_g$ = 0.72 V and (b) $V_g$ = 0.73 V. In-plane magnetic-field dependence measured in the same low-$V_g$ condition as in (a), with $V_g$ = 0.72 V, at (c) $B_{\|}$ = 1 T and (d) $B_{\|}$ = 5 T. All panels are displayed on the same logarithmic current scale, as indicated by the color bar shown on top of panel (b).

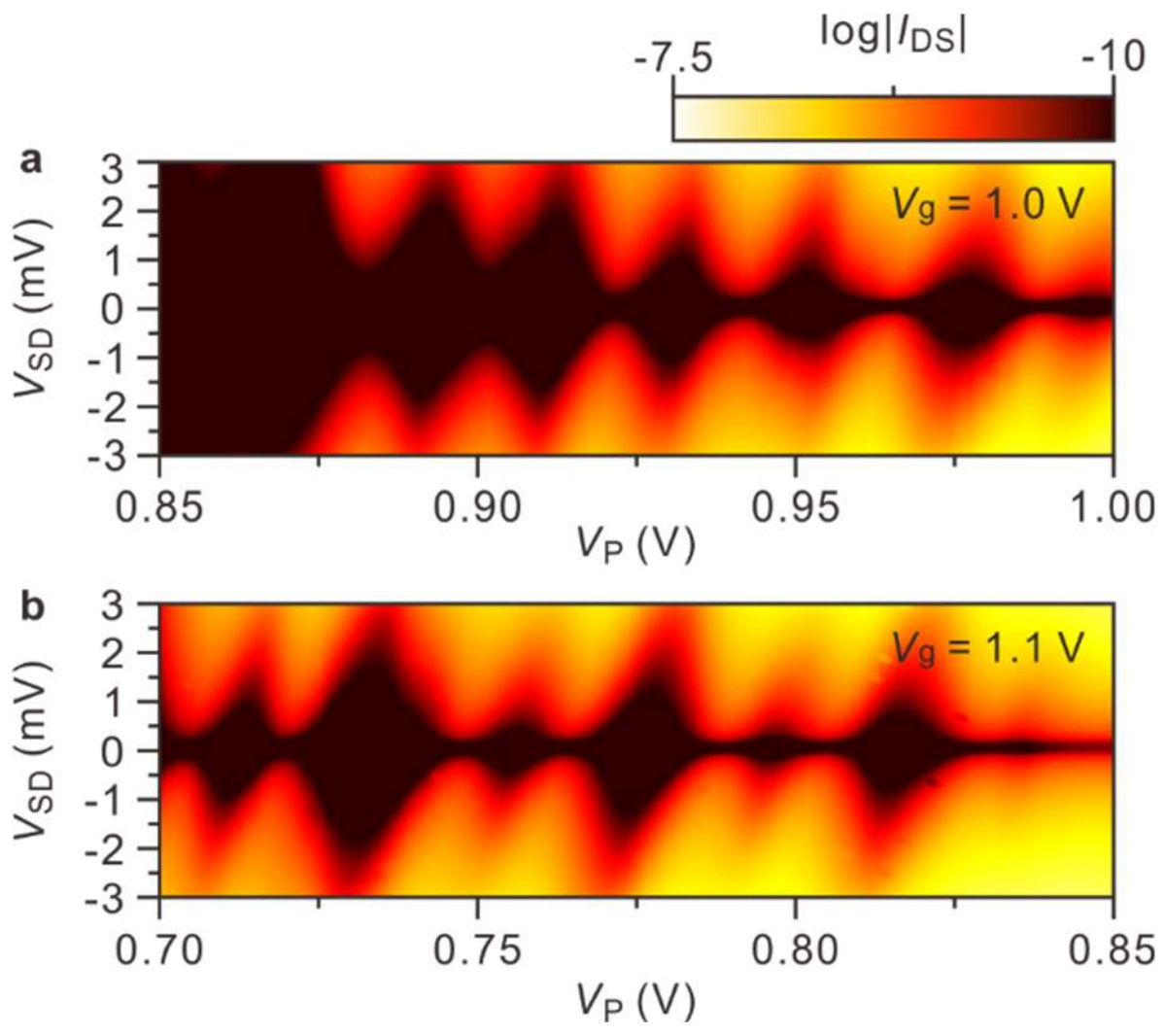

**Figure 3| Back-gate-driven change of the low-$V_g$ transport pattern.** Bias spectroscopy as a function of plunger-gate voltage $V_P$ at $V_{gg}$ = 70 V, $V_M$ = -1.0 V, and $V_L$ = $V_R$ = -1.3 V, measured at $B_{\parallel}$ = 1 T. The graphite back-gate voltage is increased from (a) $V_g$ = 1.0 V to (b) 1.1 V. All panels are displayed on the same logarithmic current scale, as indicated by the color bar. Note that the $V_P$ window differs between panels because increasing $V_g$ shifts the resonance pattern to lower $V_P$ via the back-gate–dot cross-capacitance.

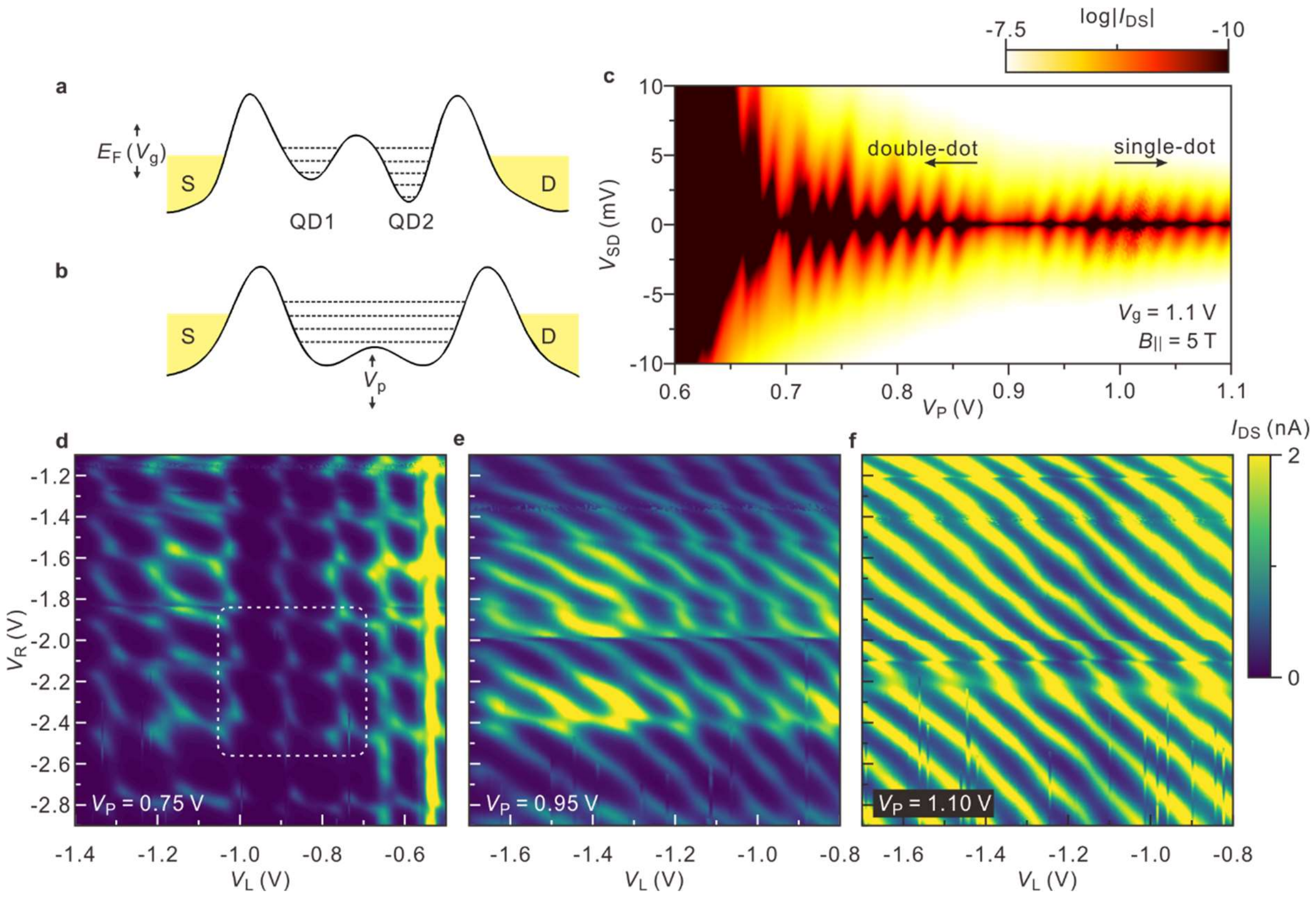

**Figure 4| Plunger-gate tuning of the higher-$V_g$ regime.** (a) Schematic asymmetric double-well potential illustrating a double-dot configuration. Here, the graphite back gate tunes the electrochemical potential, indicated as $E_F$ ($V_g$). (b) Schematic potential profile after further tuning by the plunger gate $V_P$, which lowers the central barrier and makes the two minima more merged. (c) Bias spectroscopy at $V_g$ = 1.1 V and $B_{\parallel}$ = 5 T, showing a crossover from a double-dot pattern at lower $V_P$ to a more single-dot-like pattern at higher $V_P$. Two-gate maps as a function of $V_L$ and $V_R$ at (d) $V_P$ = 0.75 V, (e) 0.95 V, and (f) 1.10 V. The color bar for panels (d-f) is placed next to panel (f).

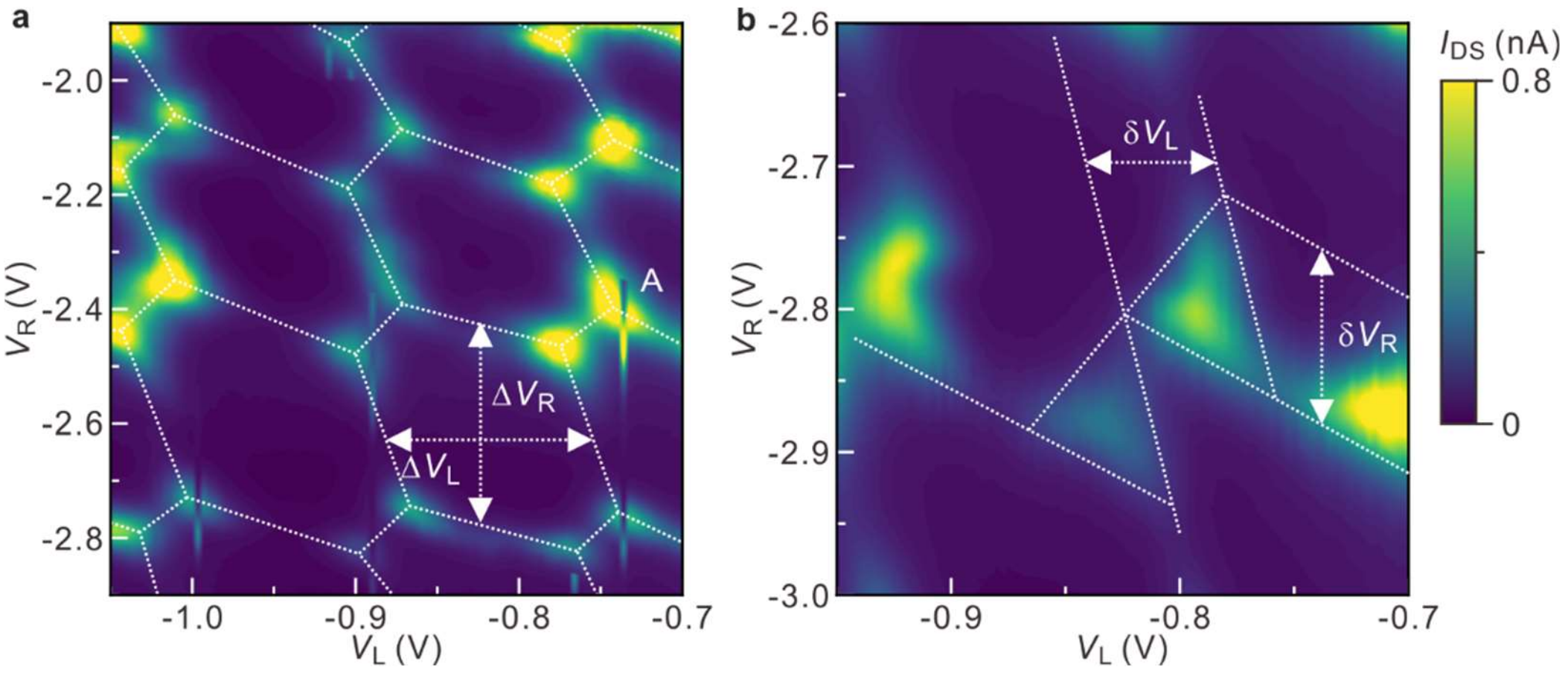

**Figure 5| Charge stability diagram of a double quantum dot.** (a) Magnified charge stability diagram of the double quantum dot, taken from the dashed rectangular region in Fig. 4(d). The honeycomb pattern characteristic of a double-dot system is clearly resolved. (b) Bias triangle measured at $V_{SD}$ = 1.5 mV at point A in (a), from which the lever arms and capacitances are extracted.